\documentclass{PoS}

\pdfoutput=1

\usepackage{amsmath,amssymb,amsfonts}
\usepackage{subfigure}

\newcommand{\be}{\begin{equation}}
\newcommand{\ee}{\end{equation}}

\title{Mass anomalous dimension and running of the coupling in SU(2) with six fundamental fermions}

\ShortTitle{$SU(2)$ with Six Fundamental Fermions}

\author{Francis Bursa\\
        Jesus College, University of Cambridge, UK\\
        E-mail: \email{fwb22@cam.ac.uk}}

\author{Luigi Del Debbio\\
        University of Edinburgh, UK\\
        E-mail: \email{luigi.del.debbio@ed.ac.uk}}

\author{Liam Keegan\\
        University of Edinburgh, UK\\
        E-mail: \email{liam.keegan@ed.ac.uk}}

\author{Claudio Pica\\
	$CP^3$-Origins and IMADA,
	University of Southern Denmark, Denmark\\
        E-mail: \email{pica@cp3.sdu.dk}}

\author{\speaker{Thomas Pickup}\\
        University of Oxford, UK\\
        E-mail: \email{pickup@thphys.ox.ac.uk}}

\abstract{We simulate SU(2) gauge theory with six massless fundamental Dirac fermions. By using the Schr\"odinger Functional method we measure the running of the coupling and the fermion mass over a wide range of length scales. We observe very slow running of the coupling and construct an estimator for the fermion mass anomalous dimension giving $0.135 <\gamma< 1.03$ in the region compatible with an IR fixed point.}

\dedicated{\flushleft{CP3-Origins-2010-43}}

\FullConference{The XXVIII International Symposium on Lattice Field Theory, Lattice2010\\
		June 14-19, 2010\\
		Villasimius, Italy}

\begin{document}

\section{Introduction}

The recent interest in phenomenologically viable Technicolor models has driven the exploration of the phase diagram of $SU(N)$ gauge theories\cite{Dietrich:2006cm}. Investigations have focused on the search for asymptotically free theories with infrared fixed points (IRFP) \cite{Banks:1981nn}, known as the conformal window. Such theories  have a non-trivial zero in the beta function which leads to conformal behaviour in the infrared. 

In particular, Walking Technicolor  eases the traditional tension between large standard model fermion masses and small flavour changing neutral currents (FCNC) by having the Technicolor gauge group run very slowly, walk, between the Technicolor and Extended Technicolor scales.   
This causes an enhancement in $\langle\bar\Psi\Psi\rangle_{ETC}$,
\be
\langle\bar\Psi\Psi\rangle_{ETC} \sim \langle\bar\Psi\Psi\rangle_{TC}\left(\frac{M_{ETC}}{\Lambda_{TC}}\right)^\gamma
\ee
where $\gamma$ is the fermion anomalous dimension. Since the standard model fermion masses are proportional to $\langle\bar\Psi\Psi\rangle_{ETC}$ this allows large fermion masses whilst retaining acceptably small FCNCs.

Specifically, this requires that the underlying gauge theory be just below the conformal window and, equally importantly, that it have a large anomalous dimension ($\gamma\approx1$). 
Currently, most of the effort has gone into investigating theories based on $SU(3)$ with large numbers of fundamental fermions or gauge groups with fermions in higher representations \cite{Dietrich:2006cm}.  

Instead, in this work, we study $SU(2)$ with fundamental fermions. This is attractive as it requires fewer fermions to be conformal than $SU(3)$, which leads to a smaller na\"ive S parameter and reduced computation time. 
Theoretical estimates for the lower end of the conformal window vary over the range $N_f\sim4-8$ \cite{Sannino:2009aw} so we decided to focus on the case of six fermions. 
 We measure the running of both the coupling and the fermion mass using the  Schr\"odinger Functional method. Unfortunately, if the running is small, it is difficult to distinguish between a truly conformal theory and one that is merely running slowly. The running of the mass is easier to determine but requires the location of a fixed point to calculate the anomalous dimension.  For examples of other theories investigated using the  Schr\"odinger Functional see e.g. \cite{Appelquist:2009ty,Bursa:2009we,DeGrand:2010na}.
 
Subsequent to this talk being given, a fuller account of our findings appeared in \cite{Bursa:2010xn} and we refer the reader to this for more detail.

\section{Formulation}
We define the running coupling $\bar g^2$ using the Schr\"odinger
Functional (SF) formalism \cite{Luscher:1991wu,Luscher:1992an}. 
Aside from the fermionic content we follow the method of \cite{Bursa:2009we}. The system
is defined on a hypercubic lattice of size $L$ with periodic boundary
conditions in the spatial directions and Dirichlet boundary conditions
in the temporal direction. The boundary spatial link matrices are set
to:
\be
U(x,k)|_{t=0}=exp[\eta\tau_3a/iL],\;\;\;\;U(x,k)|_{t=L}=exp[(\pi-\eta)\tau_3a/iL]
\ee
with $\eta=\pi/4$\cite{Luscher:1992zx}. The boundary fermionic fields obey the following 
relations:
\be
P_+\psi=0,~\overline{\psi}P_-=0\,\,~\mathrm{at}~t=0,\;\;\;\;\;\;\;\;P_-\psi=0,~\overline{\psi}P_+=0\,\,~\mathrm{at}~t=L\, 
\ee
with the projection operators defined by $P_\pm = (1/2)(1\pm\gamma_0)$.

We define the coupling constant through the relation
\be
\bar{g}^2 = k \left\langle \frac{\partial S}{\partial \eta}\right\rangle^{-1},
\ee
where  $k = -24L^2/a^2 \sin(a^2/L^2(\pi-2\eta))$ is chosen so that $\bar{g}^2=g_0^{\;2}$ to leading order in perturbation theory. This is a non-perturbative definition of the coupling that only depends on $L$ and the lattice spacing $a$. We can then remove the dependence on $a$ by taking the continuum limit.

\begin{figure}
\centering
\includegraphics[width=0.7\textwidth]{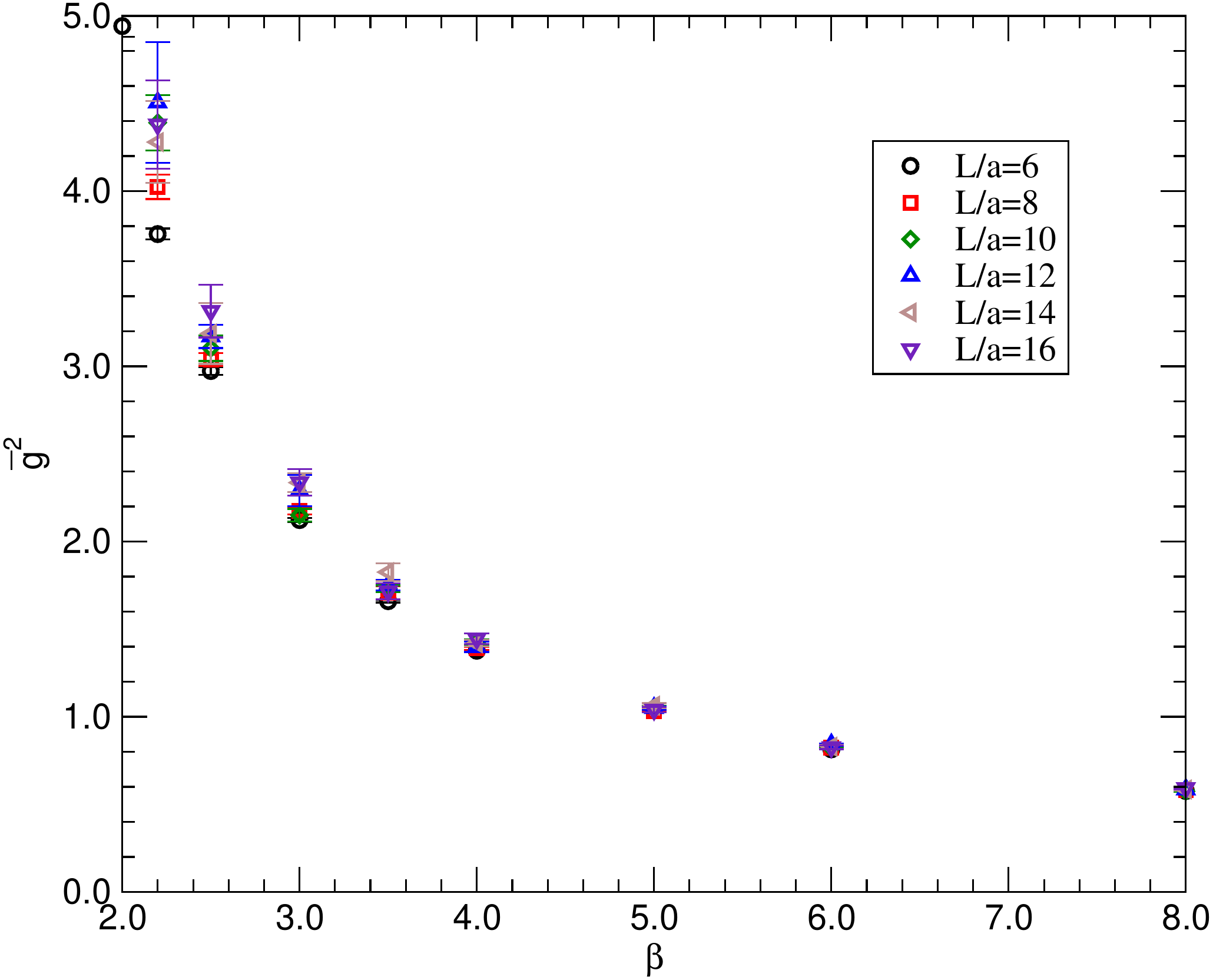}
\caption{Data for the running coupling in the Schr\"odinger Functional Scheme. Simulations are performed over a range of bare couplings $\beta$ and lattice sizes $L/a$.} 
\label{fig:coupling_data}
\end{figure}

We use the Wilson plaquette gauge action, together with fundamental Wilson fermions
and an RHMC algorithm with four pseudofermions \cite{DelDebbio:2008zf}. SF boundary conditions  allow us to simulate at $\kappa_c$, which we define as the value of $\kappa$ at which the PCAC mass vanishes. 
\be
m_{PCAC} = \frac{\frac{1}{2}(\partial_0+\partial_0^*)f_A(L/2)}{2f_P(L/2)}
\ee
where $f_A$ and $f_P$ are the axial and pseudoscalar correlation functions and $\partial_0,\partial_0^*$  are the forwards and backwards lattice derivatives respectively.

As in \cite{Capitani:1998mq}, we use $f_P$ to define the pseudoscalar density renormalisation constant:
\be
Z_P(L) =\sqrt{3f_1}/f_P(L/2)
\ee
where $f_1$ is a purely boundary term. By observing the scale dependence of $Z_P$ we can calculate the mass anomalous dimension. However, in order to obtain the correct normalisation this has to be measured on lattices with the spatial link matrices at $t = 0$ and $t = L$ set to unity.

We perform measurements on lattices of  size $L=6,8,10,12,14,16$ for a range of bare couplings between $\beta=2$ and $\beta=8$.  Measurements to determine $\kappa_c$ are made on lattices with unit boundary conditions for $L\leq12$ and extrapolated from these for $L=14,16$. Full details of the bare parameters used can be found in \cite{Bursa:2010xn}.

\section{Running Coupling}

As shown in Fig.~\ref{fig:coupling_data}, we measure the SF coupling $\bar{g}^2$ for a range of $\beta,L$. It is clear that there is little variation with $L/a$, indicating that the coupling runs very slowly.  In order to quantify this we calculate the lattice step-scaling function:
\be
\Sigma(u,s,a/L)= \bar{g}^2(g_0,sL/a)\vert_{\bar{g}^2(g_0,L/a)=u}.
\ee
Then using data from multiple values of $L/a$ we can calculate its continuum limit:
\be
\sigma(u,s)= \lim_{a/L\rightarrow 0}\Sigma(u,s,a/L)
\ee
Throughout this work we set $s=3/2$. Using our measured results we interpolate in $a/L$  at each $\beta$ to find $\bar{g}^2(g_0,sL/a)$ at $L = 9\frac{1}{3},10\frac{2}{3},15$. The results for $L=6$ were found to contain large artifacts and are not included in the interpolation or further analysis. For each $L$ we then interpolate in $\beta$ using the functional form \cite{Appelquist:2009ty}:
\be
\frac{1}{\bar{g}^2(\beta,L/a)}=\frac{\beta}{2N}\left [ \sum^n_{i=0}c_i\left ( \frac{2N}{\beta}\right )^i \right ].
\label{eq:fit_sigma}
\ee
See \cite{Bursa:2010xn} for full details on the optimal parameters and associated errors.
\begin{figure}
\centering
\subfigure[Constant Extrapolation]{
\includegraphics[width=0.45\textwidth]{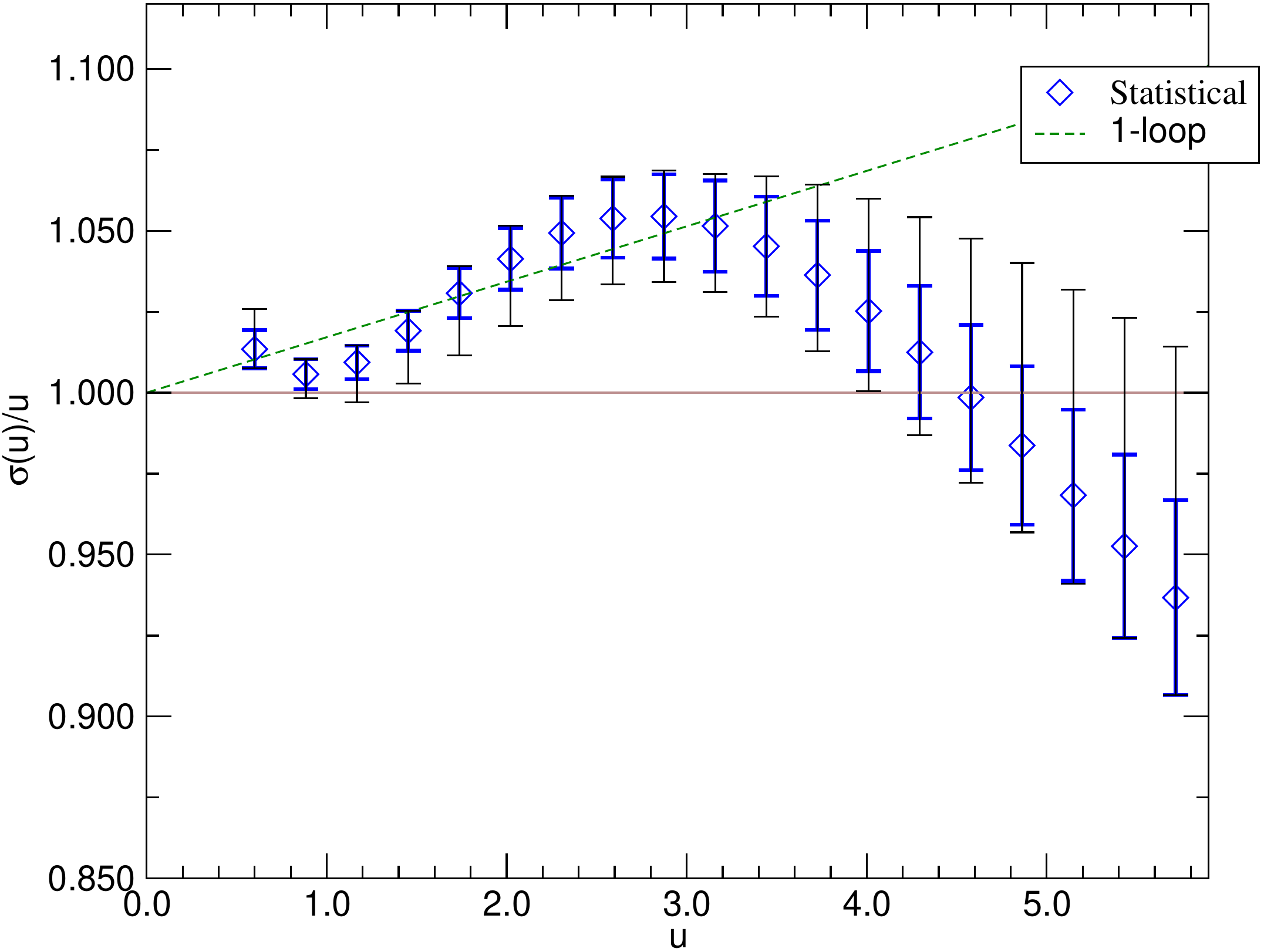}
\label{fig:const_sigma}
}
\subfigure[Linear Extrapolation]{
\includegraphics[width=0.45\textwidth]{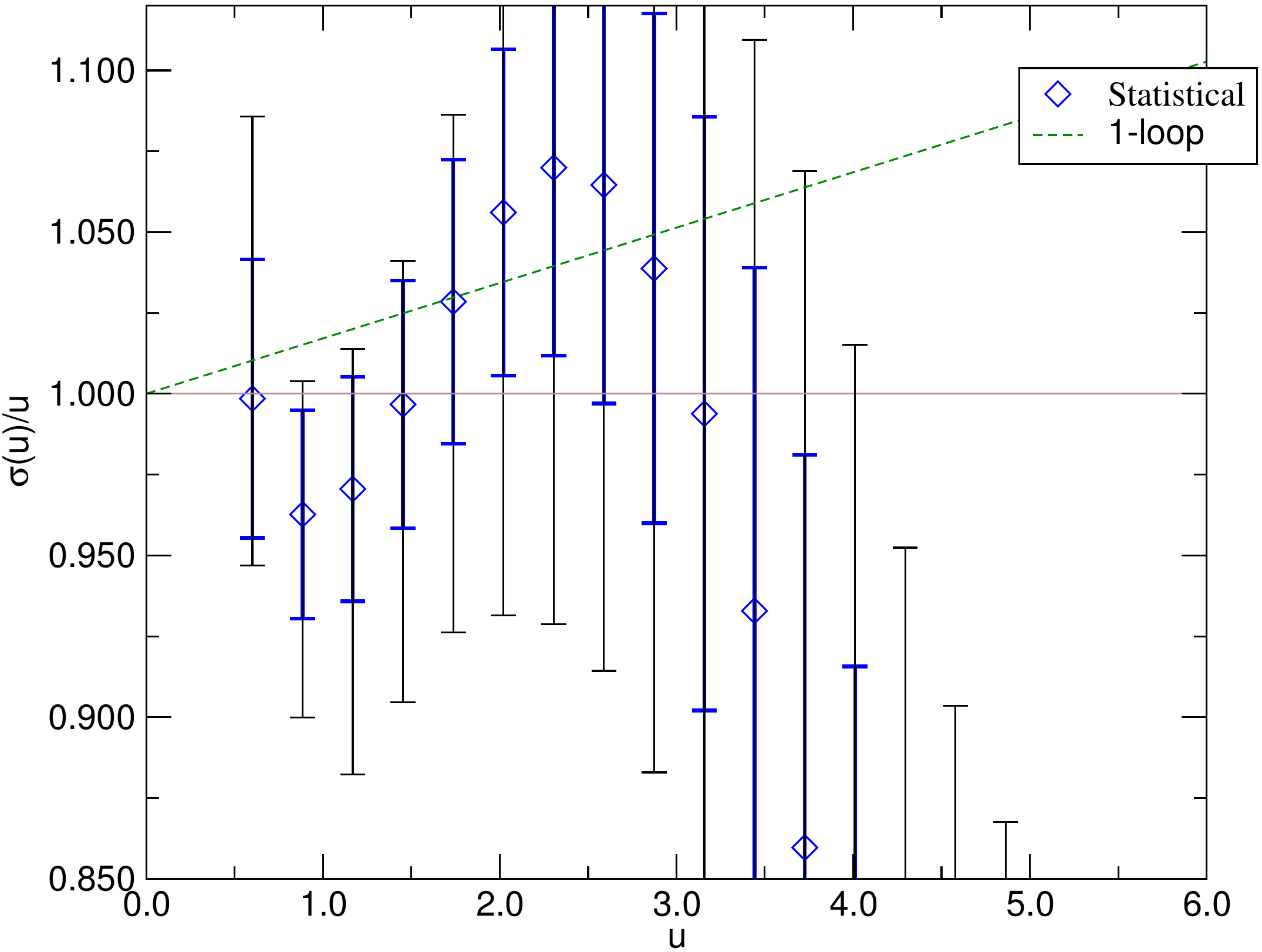}
\label{fig:lin_sigma}
}
\caption{Results for $\sigma(u)/u$ using both constant and linear continuum extrapolations. Blue error bars indicate statistical error, black error bars include systematic errors. A zero in the beta function is indicated by $\sigma(u)/u=1$.} 
\label{fig:sigma_results}
\end{figure}

We are now able to calculate $\Sigma(u,3/2,a/L)$ using the fits from Eq.~\ref{eq:fit_sigma} for $L=8,9\frac{1}{3},10,10\frac{2}{3}$. We can use these to calculate the continuum value $\sigma(u)\equiv\sigma(u,3/2)$. We perform both a constant continuum extrapolation, using the two values of $a/L$ closest to the continuum limit, and a linear continuum extrapolation. The results for the constant extrapolation are shown in Fig.~\ref{fig:const_sigma} and for the linear extrapolation in Fig.~\ref{fig:lin_sigma}. The figures show $\sigma(u)/u$ with the blue error bars giving the purely statistical error and the black error bars including the systematic errors.

The errors in the linear extrapolation are particularly large and indicate that our data cannot provide good constraints on the fits. The results from the
constant continuum extrapolation are consistent with a fixed point
with $\bar{g}^2>4.02$. However, at our strongest couplings we are
still unable to rule out the possibility that no fixed point is observed. On the other hand, it is clear that at large $\bar{g}^2$ $\sigma(u)$ is considerably below the 1-loop prediction. Further data is required in order to decrease the errors on the linear extrapolation and thus fully account for the errors in taking the continuum limit.

\begin{figure}
\centering
\includegraphics[width=0.7\textwidth]{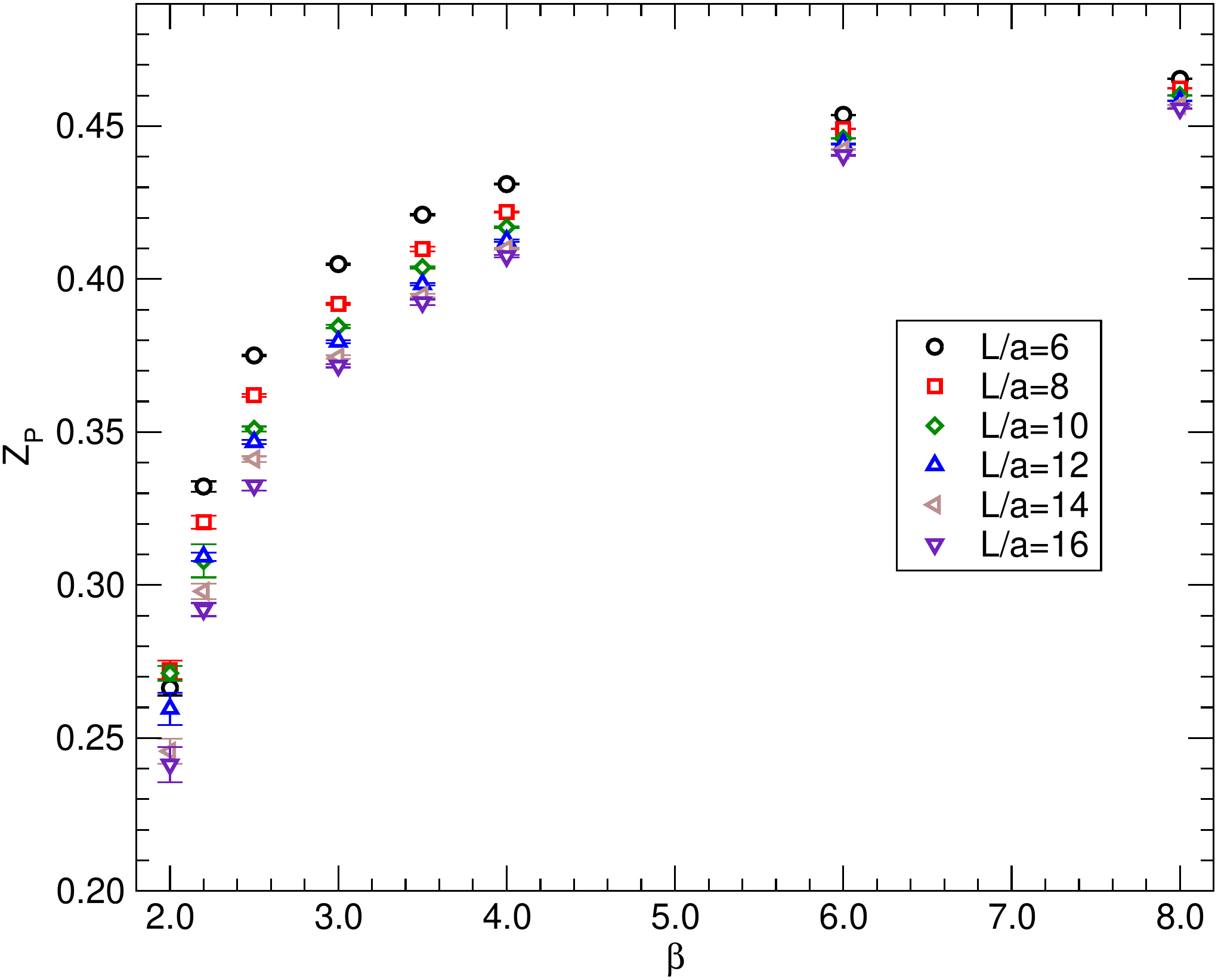}
\caption{Data for the pseudoscalar density renormalisation constant $Z_P$ computed in the Schr\"odinger functional scheme.} 
\label{fig:ZP_data}
\end{figure}

\begin{figure}
\centering
\subfigure[]{
\includegraphics[width=0.45\textwidth]{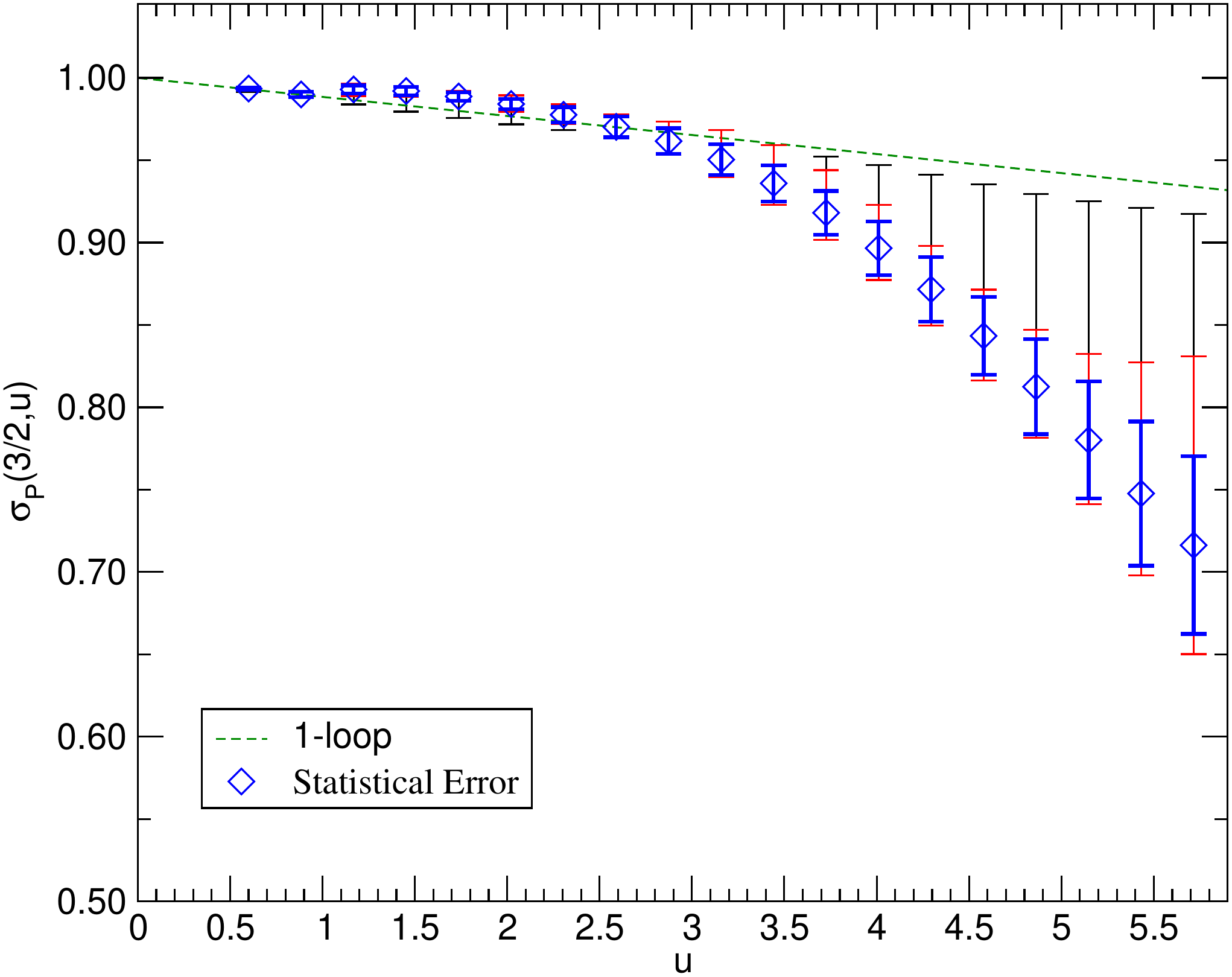}
\label{fig:sigma_P}
}
\subfigure[]{
\includegraphics[width=0.45\textwidth]{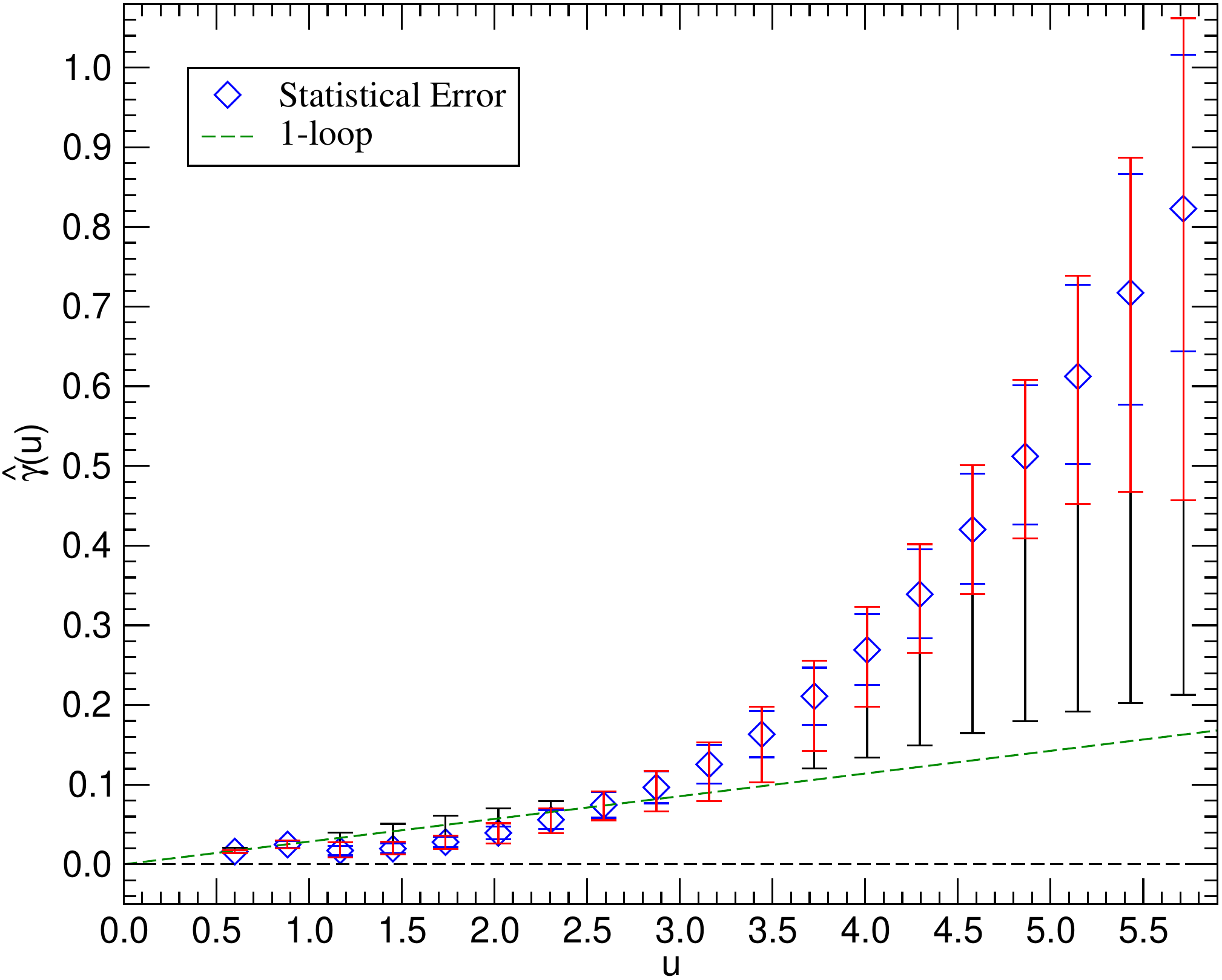}
\label{fig:gamma}
}
\caption{Results for $\sigma_P(u)$are shown in \protect \subref{fig:sigma_P} and the corresponding results for $\hat\gamma(u)$ are shown in \protect \subref{fig:gamma}. The blue error bars correspond to  the statistical error, the black error bars include all systematic errors and the red error bars do not include the continuum extrapolation} 
\label{fig:sigma_P_results}
\end{figure}

\section{Mass Anomalous Dimension}
We measure the pseudoscalar density renormalisation constant $Z_P$ over a a range of $\beta, L$; our results are shown in Fig.~\ref{fig:ZP_data}. Unlike the coupling data in Fig.~\ref{fig:coupling_data}, there is noticeable variation with $L/a$ at fixed $\beta$, indicating a non-zero anomalous dimension $\gamma$.

We extract $\gamma$ from the step-scaling function for the mass. We define the lattice version as
\be
\Sigma_P(u,s,a/L)=\left . \frac{Z_P(g_0,sL/a)}{Z_P(g_0,L/a)}\right |_{\bar{g}^2(g_0,L/a)=u}.
\label{eq:Sigma_P}
\end{equation}
From this we can define the continuum mass step-scaling function,
\be
\sigma_P(u,s)=\lim_{a/L\rightarrow0}\Sigma_P(u,s,a/L).
\label{eq:sigma_P}
\end{equation}
Again, we set $s=3/2$ throughout our calculations and discard the $L=6$ results. We interpolate in $L$ at fixed $\beta$ to obtain results for $Z_P$ at $L = 9\frac{1}{3},10\frac{2}{3},15$. Then for each $L$ we interpolate in $\beta$ with the functional form \cite{Bursa:2009we}:
\be
Z_P(\beta,L/a)=\sum_{i=0}^n c_i \left( \frac{1}{\beta}\right )^i.
\label{eq:Z_P_fit}
\ee
Full details on the best fit parameters and associated error can be found in \cite{Bursa:2010xn}.

We calculate $\Sigma_P(u,3/2,a/L)$ using the fits from Eq.~\ref{eq:Z_P_fit} in Eq.~\ref{eq:Sigma_P} and then perform the continuum extrapolation to recover $\sigma_P(u)\equiv\sigma_P(u,3/2)$. Unlike for the coupling, the errors on both linear and constant continuum extrapolation are sufficiently small to allow us to combine them to estimate the systematic error due to taking the  continuum limit. In Fig.~\ref{fig:sigma_P} we plot $\sigma_P(u)$ including systematic errors from both the fitting procedure and taking the continuum limit.

When the coupling is weakly running we can construct an estimator for the anomalous dimension, $\hat\gamma$, from $\sigma_P(u)$ , 
\be
\hat\gamma(u)=-\frac{\log\left|\sigma_P(u,s)\right|}{\log\left| s\right|}
\ee
This estimator becomes exact at a fixed point
\cite{Bursa:2009we}. Note also that $\gamma$ itself is only a scheme
independent quantity at a fixed point. 

We plot the estimator in Fig.~\ref{fig:gamma} using the same error bars as for $\sigma_P(u)$. We see that at small values of $u$ it closely follows the 1-loop prediction. At stronger coupling both the value and error increases and at our strongest couplings becomes compatible with $\gamma\approx1$ as desired for walking technicolor. However, at the lower end of our prediction for a fixed point, $u=4.02$, $\gamma$ could be as small as $0.135$.  

\section{Outlook}

In these proceedings we have presented results for the running of the coupling $\bar{g}^2$ and the mass anomalous dimension $\gamma$ in the Schr\"odinger  Functional scheme for $SU(2)$ with six fundamental fermions.

We find that our results are consistent with a fixed point at $\bar{g}^2>4.02$. However, at our strongest couplings ($\bar{g}^2\approx5.5$) we are unable to rule out that the beta function remains negative and that there is no fixed point. In either case, we observe that $\bar{g}^2$ runs very slowly across the range investigated. Currently, the errors on the linear continuum extrapolation are too large to allow us to calculate our systematics due to taking the continuum limit. Further computation is required to address this.

Determination of the anomalous dimension $\gamma$ is less affected by systematic errors. The primary source of error is due to uncertainty over the running of $\bar{g}^2$. Assuming that we do observe a fixed point then we find that $0.135<\gamma<1.03$, with the, phenomenologically  interesting, upper range only being favoured at the extreme range of the coupling investigated. At weaker couplings our results agree well with 1-loop perturbation theory. However, at stronger couplings they favour a larger anomalous dimension.

Currently, the greatest source of uncertainty is due to being unable to perform a linear continuum extrapolation for $\bar{g}^2$. In order to address this, we are currently performing simulations at larger lattice sizes. 


\bibliographystyle{tp}
\bibliography{../../../../tp.bib}

\end{document}